# High space-bandwidth in quantitative phase imaging using partially spatially coherent optical coherence microscopy and deep neural network

Ankit Butola[1, 2#], Sheetal Raosaheb Kanade[1#], Sunil Bhatt[1], Vishesh Kumar Dubey[2], Anand Kumar[1], Azeem Ahmad[2], Dilip K Prasad[3], Paramasivam Senthilkumaran[4], Balpreet Singh Ahluwalia[2,5] and Dalip Singh Mehta[1]

[1] Bio-photonics Laboratory, Department of Physics, Indian Institute of Technology Delhi, Hauz-Khas, New Delhi- 110016, India.
[2] Department of Physics and Technology, UiT The Arctic University of Norway, Tromsø, Norway.
[3] Department of Computer Science, UiT The Arctic University of Norway, Tromsø, Norway.
[4] Department of Physics, Indian Institute of Technology Delhi, Hauz-Khas, New Delhi- 110016, India.
[5] Department of Clinical Science, Intervention and Technology Karolinska Institutet, Stockholm, Sweden.
[#] These authors contributed equally in this work.
Corresponding author: ankitbutola321@gmail.com

**Abstract**

Quantitative phase microscopy (QPM) is a label-free technique that enables to monitor morphological changes at subcellular level. The performance of the QPM system in terms of spatial sensitivity and resolution depends on the coherence properties of the light source and the numerical aperture (NA) of objective lenses. Here, we propose high space-bandwidth QPM using partially spatially coherent optical coherence microscopy (PSC-OCM) assisted with deep neural network. The PSC source synthesized to improve the spatial sensitivity of the reconstructed phase map from the interferometric images. Further, compatible generative adversarial network (GAN) is used and trained with paired low-resolution (LR) and high-resolution (HR) datasets acquired from PSC-OCM system. The training of the network is performed on two different types of samples i.e. mostly homogenous human red blood cells (RBC) and on highly heterogenous macrophages. The performance is evaluated by predicting the HR images from the datasets captured with low NA lens and compared with the actual HR phase images. An improvement of 9× in space-bandwidth product is demonstrated for both RBC and macrophages datasets. We believe that the PSC-OCM+GAN approach would be applicable in single-shot label free tissue imaging, disease classification and other high-resolution tomography applications by utilizing the longitudinal spatial coherence properties of the light source.

## 1. Introduction

Quantitative phase imaging (QPI) is an emerging label-free technique to visualize sub-micron changes in various cells and tissues. QPI measures the path length shift associated to specimen which contain the information about both refractive index and local thickness of the structure [1]. The most common approach to extract the path length shift i.e. phase information of the sample is based on the principle of holography[2]. In holography, object information is encoded in the form of spatially modulated signal due to the interference between reference and sample field which can be extracted using different reconstruction algorithm. Since it was first introduced two decades ago[2], QPI has been developed gradually (both experimentally and computationally) to improve the acquisition speed, space-bandwidth product (SBP), spatial and temporal phase sensitivity [3-7]. For example, Fourier ptychography microscopy (FPM) has been shown to improve the SBP in QPI but required number of measurements to achieve high resolution image from low numerical aperture (NA) lens [5]. Recently, deep learning (DL) based FPM is proposed to avoid the need of number of measurements and required less frame to achieve higher resolution[8].

On the other hand, common path coherent QPI techniques such as in-line holography [9], diffraction phase microscopy (DPM) [3], QPI-unit (QPIU) [10] and lateral shearing interferometric microscopy [11] are used to improve the temporal stability of the phase microscopy system. These techniques have been also integrated with deep learning to improve the space-bandwidth, automated classification of human red blood cells (RBC), spermatozoa, anthrax spores and among others [12-15]. However, the reconstruction algorithm associated with aforementioned techniques suffers with pixel limited resolution[9], twin image problem[10, 11] and poor spatial phase sensitivity[16] thus cannot offer the fine structural information over the large field of view (FOV) of the specimens. Off-axis incoherent QPI i.e. using broadband source such as halogen lamp and LED can be used to overcome aforementioned problems but required multiple frame to extract the phase information due to poor temporal coherence [17-19].

Moreover, the partially spatially coherent (PSC) source has been used to bridge the single-shot, high spatial phase sensitivity gap between coherent and incoherent phase imaging techniques. The longitudinal spatial coherence properties of the PSC source has been utilized in the previous studies for various applications such as surface profilometry[20, 21], full-field optical coherence tomography (FF-OCT) [22, 23], holography [24], spatial coherence tomography [25], and others [4, 26]. Additionally, the longitudinal spatial coherence properties of the PSC source can be utilized for high resolution sectioning in FF-OCT [22]. Despite so many applications, the spatial resolution and the outcome of the aforementioned techniques is limited with the NA of the microscopic objective lens and hence cannot offer single shot large FOV imaging of the specimens. Therefore, a technique which can offer fast, accurate and large FOV imaging with PSC source would be more applicable in noise free topography and tomography applications of industrial and biological specimens.

Here, we propose a PSC-optical coherence microscopy (OCM) assisted with deep neural network (DNN) to achieve both high resolution and spatial phase sensitivity in diverse biological cells. The PSC source is synthesized by passing the direct laser through rotating diffuser and multi-multimode fiber bundle (MMFB). The output of the MMFB act as a temporally coherent and spatially incoherent source which coupled at input port of Linnik type interferometer. The interferometric images are acquired with both low and high NA lenses to further extract phase map of the object. The low-resolution phase map transformed into high resolution in a single feedforward step using generative adversarial network (GAN). The network first trained and optimized to convert low resolution images into high resolution images using large training datasets. After sufficient training, the performance of the network is demonstrated on two biological cells i.e. human RBC and macrophages. The generated images of both RBC and macrophages are further compared with the ground truth (high resolution image) to evaluate the performance of the network.

## 2. Material and methods

### 2.1. Experimental details

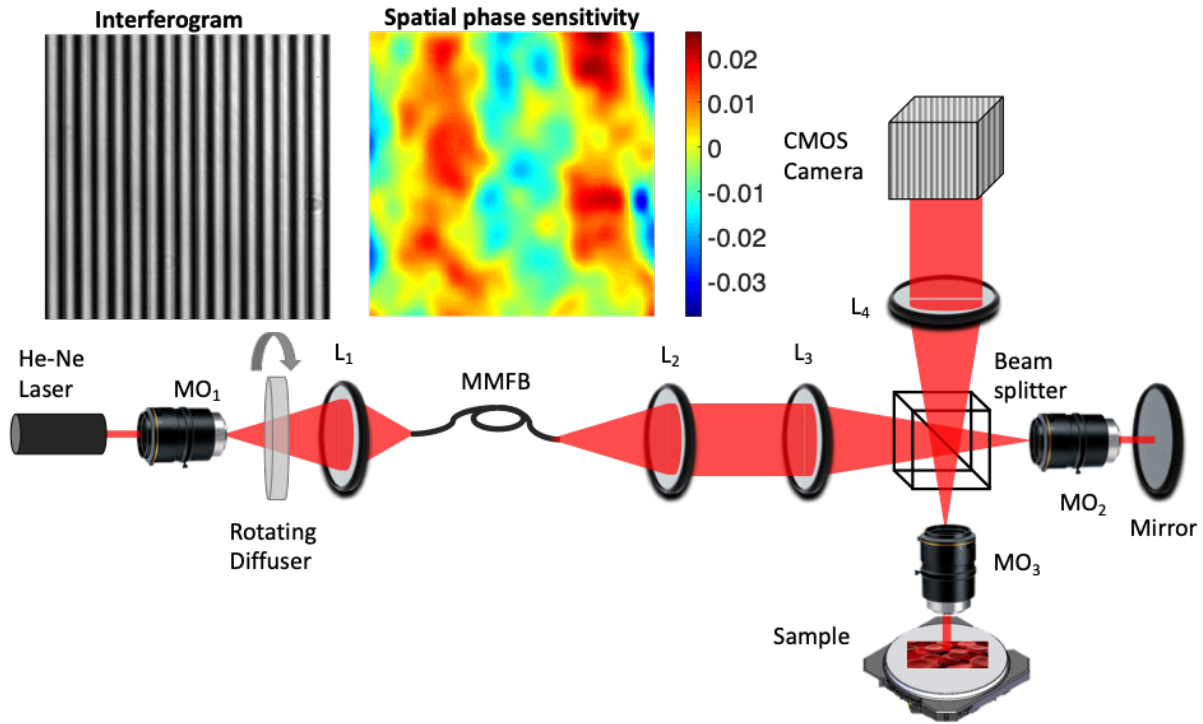

Figure 1: Schematic diagram of partially spatially coherent optical coherence microscopy (PSC-OCM) system. The PSC source synthesized by introducing spatial and temporal diversity in the path of laser beam to reduce the spatial coherence and thus average out speckle pattern in the final image. The spatial and temporal diversity generated by using rotating diffuser and multi-multimode fiber bundle (MMFB). The interference pattern shows the significant improvement in the spatial phase sensitivity of the system. MO: Microscopic objective lens; L: Lens.

Figure 1 depicts the schematic diagram of PSC-OCM system. A PSC source is synthesized by passing a monochromatic light source through a microscopic objective lens (MO$_1$) and a rotating diffuser. A rotating diffuser scattered the input photons into multiple direction which further focused by a lens L$_1$. The focused light coupled at the input port of multi-multimode fiber bundle (MMFB). The rotating diffuser and MMFB is used to generate spatial and temporal diversity in the path of laser beam. It has been shown previously that the effect of spatial, temporal and angular diversity can be used to significantly reduce the speckle pattern of the coherent light source [27]. In other words, average the incoming photons by performing spatial and temporal diversity reduce the spatial coherence of the light source and further improve the quality of final image. The output of MMFB coupled into Linnik type interferometer. At the input port of interferometer, a PSC light beam is first collimated and then focused into back focal plane of MO$_2$ (reference arm) and MO$_3$ (sample arm). The focused light passes through the objective lens to illuminate the sample. The back reflected light in the sample arm contain the sample information and interfere with the reference beam at beam splitter plane. The interference signal further collimated and projected into camera plane using tube lens L$_4$. The reference beam can be approximated by a plane wave and is tilted to generate the high fringe density at the camera plane. The intensity distribution at the camera plane can be written mathematically:

$$I(x,y) = a(x,y) + b(x,y)\cos[2\pi i(f_x x + f_y y) + \phi(x,y)] \quad (1)$$

$a(x,y)$ and $b(x,y)$ are the background (DC) and the modulation terms, respectively. Spatially varying phase $\phi(x,y)$ contains information about the specimen and $f_x$, $f_y$ are the spatial carrier frequencies of interferogram. The intensity modulation expressed as:

$$I(x,y) = a(x,y) + c(x,y)\exp[2\pi i(f_x x + f_y y)] + c^*(x,y)\exp[-2\pi i(f_x x + f_y y)] \quad (2)$$

where

$$c(x,y) = B(x,y)\exp(i\phi(x,y)) \quad (3)$$

The Fourier transform[28] of Eq. 2 can be written as:

$$FI(\xi_x, \xi_y) = Fa(\xi_x, \xi_y) + Fc(\xi_x - f_x, \xi_y - f_y) + Fc^*(\xi_x + f_x, \xi_y + f_y) \quad (4)$$

The term $Fa(\xi_x, \xi_y)$ represents the background (DC) term at the origin in the Fourier plane and $Fc(\xi_x - f_x, \xi_y - f_y)$ corresponds to +1 order term contains information about the object at $(+f_x, +f_y)$ position. Similarly, $Fc^*(\xi_x + f_x, \xi_y + f_y)$ is –1 order at position $(-f_x, -f_y)$ carry complex conjugate information. After applying Fourier filtering of zero and – 1 order terms, Eq. 4 reduced into the following:

$$FI(\xi_x, \xi_y) = Fc(\xi_x - f_x, \xi_y - f_y) \quad (5)$$

The filtered spectrum shifted at the origin and then inverse Fourier transformed to retrieve the complex signal $c(x,y)$, subsequently the wrapped phase map from the following expression:

$$\phi(x,y) = \tan^{-1}\left[\frac{Im(c(x,y))}{Re(c(x,y))}\right] \quad (6)$$

where Im and Re are the imaginary and real part of the complex signal. The reconstructed wrapped phase map lies between $-\pi$ to $+\pi$. The unwrapping of the phase map is done by using transport of intensity (TIE) equations[29]. It is found that TIE unwrapping is faster and perform better for thicker sample as compare to the other unwrapping algorithm[30]. In TIE bases phase unwrapping, first a complex field is created with constant amplitude and phase $\phi(x,y)$, given by

$$c_o(x,y;0) = \exp(i\phi(x,y)) \quad (7)$$

The field $c_o(x,y;0)$ propagated to closely spaced plane to obtain the longitudinal intensity derivative, Mathematically

$$c_o(x,y;\pm z) = c_o(x,y;0) * s(x,y;\pm z) \quad (8)$$

The Eq. (8) shows the convolution between complex field and impulse response for free-space propagation (first Rayleigh-Sommerfeld solution) where $s(x, y; z)$ represents the impulse response and can be expressed as:

$$s(x, y; z) = \frac{e^{ikR}}{2\pi R}\left(-ik + \frac{1}{R}\right)\frac{z}{R} \tag{9}$$

The impulse response can be correlate with the angular spectrum transfer function for free space propagation using following expression:

$$S(f_x, f_y; z) = F[s(x, y; z)] = \exp\left[iz\sqrt{k^2 - 4\pi^2(f_x^2 + f_x^2)}\right] \tag{10}$$

The longitudinal intensity derivative can be estimate using central difference relation:

$$\frac{\partial I}{\partial z} = \frac{|c_o(x, y; \Delta z)|^2 - |c_o(x, y; -\Delta z)|^2}{2\Delta z} \tag{11}$$

Finally, applying inverse Laplacian transform, the unwrapped phase can be shown as

$$\phi(x, y; z) = \frac{-k}{I}\nabla_{x,y}^{-2}\left(\frac{\partial I}{\partial z}\right) \tag{12}$$

The Eq. (12) shows the unwrapped phase and since implemented using the FFT algorithm, the timing performance is better than any other method. The interferogram and the spatial phase sensitivity of the PSC-OCM system can be seen in Fig 1. Spatial noise presents in the system can be characterize by measuring the spatial phase sensitivity of the system. It is calculated by capturing an interferometric image on top of a flat mirror of surface flatness $\lambda/10$. Ideally the reconstructed phase map should be zero, but it was found ±20 mrad due to the spatial noise in the system.

## 2.2. Workflow of the framework: sample preparation, data acquisition and image preprocessing

The workflow of the proposed PSC-OCM+DNN framework for predicting high resolution phase image is shown in Fig. 2. The workflow is divided into three parts i.e. data acquisition, image registration for training of the network and finally the prediction of high resolution from low resolution image. To prepare the sample, macrophages cell lines (RAW 264.7) is used in present experiments and cell culture was carried out at UiT-The Arctic University of Norway. Cell lines cultured in a humidified atmosphere of 95% air and 5% $CO_2$ (at 37 °C) with glutamine containing RPMI-1640 medium supplemented with 10% fetal bovine serum and antibiotics (penicillin and streptomycin). To perform the experiment, cells were seeded in PDMS chambers located on reflecting silicon slides. The RBC samples are prepared by mixing with phosphate buffered saline (PBS) solutions and centrifuged for 10-15 min to isolate the RBCs from other components. Isolated RBC sample was pipetted into a polydimethylsiloxane (PDMS) chamber which was prepared on top of the reflecting silicon slides.

In data acquisition process, the raw interferograms of RBC sample is acquired using the PSC-OCM system by employing two microscopic objectives ($MO_1$: 10X, 0.25 NA and $MO_2$: 50X, 0.75 NA). The macrophages datasets are acquired by using 20X, 0.40 NA and 60X 1.20 NA objective lens. Table 1 shows the detail comparison of the SBP while acquiring data using both low NA and high NA objective lens. The low-resolution (LR) and high-resolution (HR) interferometric images are reconstructed using Fourier transform (FT) algorithm and TIE unwrapping as shown in experimental details. The LR image of RBC datasets supports approximately 25X larger field-of-view (FOV) as compared to the HR image but with approximately 1/3rd lateral resolution supported by the HR image. For macrophages datasets, the LR images are approximately 9X FOV and 1/3rd resolution compared to the HR image. Interestingly, higher depth of field of low NA lens help to preserve finer feature which might be out of focus due to high NA objective lens. To precise match LR with HR image, LR image was cropped roughly as per the approximate matching area to that of HR counterpart. The LR image was then bicubically up-sampled 5X times in both X and Y directions for RBC and 3X times for macrophages to match the pixel size of HR image. The HR and up-sampled LR image were finely registered by using normalized cross-correlation based image registration algorithm.

The cross-correlation function correlates the spatial patterns and rectifies the shifts and rotational misalignments between the images. The cross-correlation in spatial domain is given by the equation:

$$\gamma(u,v) = \frac{\sum_{x,y}[HR(x,y) - \overline{HR}_{u,v}][LR(x-u, y-v) - \overline{LR}]}{\sqrt{\sum_{x,y}[HR(x,y) - \overline{HR}_{u,v}]^2 \sum_{x,y}[LR(x-u, y-v) - \overline{LR}]^2}} \qquad (13)$$

where $\gamma(u,v)$ represents the correlation coefficient, $HR(x,y)$ (512 x 512 pixel) and $LR(x-u, y-v)$ corresponds to high-resolution and up-sampled low-resolution image, respectively. The registered pair of LR and HR image further split into training and testing set. Standard data augmentation process is used in each pair of images i.e. rotated by $\frac{\pi}{3}$ rad followed by random flipping and random jittering. The paired augmented dataset finally used to train the network. After training the network, the LR image of size 512 x 512 pixel is given as an input to the trained generator model and corresponding HR image is instantly obtained as an output in a single step. The network was programmed in Python (version 3.7.0) and implemented using TensorFlow (version 2.0) and Keras (version 2.2.0) library functions on google Colab platform. The Google hardware used for training and testing consisted of two 12GB NVIDIA Tesla K80 GPUs, an Intel Xeon CPU @ 2.20GHz and 13 GB of RAM. More details of the network are shown in the next section 2.3.

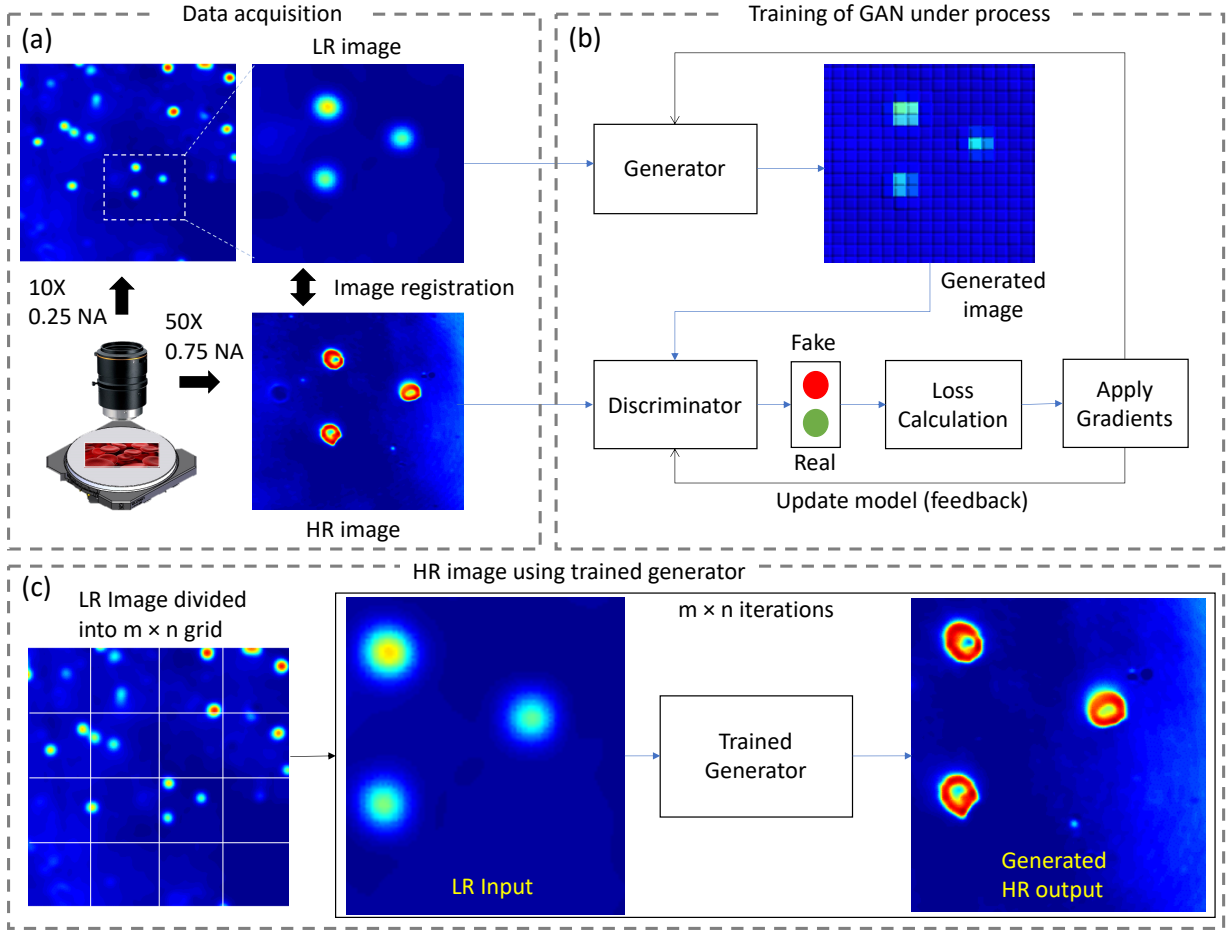

Figure 2: Workflow to achieve high-space bandwidth phase imaging in partially spatially coherent optical coherence microscopy (PSC-OCM) using generative adversarial network: (a) Reconstructed phase images using optical coherence microscopy interferograms acquired by MO 10X, 0.25 NA and 50X, 0.75 NA respectively (b) Training workflow of generative adversarial network (c) HR image with high FOV using trained generator.

Table 1: The data acquisition details in RBC, macrophages datasets and the enhancement in space-bandwidth product in the PSC-DHM+DNN framework is shown here.

|  | SBP in RBC images | SBP in Macrophages images |
|---|---|---|
| Low resolution dataset | $SBP_{10\times} = \frac{FOV_{10\times}}{0.25 \times \delta^2}$ $\left(\delta = \frac{\lambda}{2 \times NA_{10\times}}, NA_{10\times} = 0.25\right)$ | $SBP_{20\times} = \frac{FOV_{20\times}}{0.25 \times \delta^2}$ $\left(\delta = \frac{\lambda}{2 \times NA_{20\times}}, NA_{20\times} = 0.40\right)$ |
| High resolution dataset | $SBP_{50\times} = \frac{FOV_{50\times}}{0.25 \times \delta^2}$ $\left(\delta = \frac{\lambda}{2 \times NA_{50\times}}, NA_{50\times} = 0.75\right)$ | $SBP_{60\times} = \frac{FOV_{60\times}}{0.25 \times \delta^2}$ $\left(\delta = \frac{\lambda}{2 \times NA_{60\times}}, NA_{60\times} = 1.2\right)$ |
| Proposed PSC-OCM + DNN framework | $SBP_{PSC-OCM+DNN} = \frac{FOV_{10\times}}{0.25 \times \delta^2} = 9 \times SBP_{10\times}$ $\left(\delta = \frac{\lambda}{2 \times NA_{50\times}}\right)$ | $SBP_{PSC-OCM+DNN} = \frac{FOV_{20\times}}{0.25 \times \delta^2} = 9 \times SBP_{20\times}$ $\left(\delta = \frac{\lambda}{2 \times NA_{60\times}}\right)$ |

### 2.3 Deep neural network architecture:

The HR phase images are predicted from the LR datasets by using generative-adversarial network (GAN). GAN is a type of DNN consists of two building blocks, generator and discriminator as shown in Fig. 3. Throughout the training, the generator (*G*) learns the best mapping from LR image and random noise vector to HR image. The discriminator (*D*) is trained to outperform the generator by distinguishing between the mapped image by the generator and the corresponding ground truth (HR image). Initially, the LR image is given as an input to the untrained generator model. The generated image *G(LR)* and the ground truth (HR) is then fed into the discriminator. The calculated losses based on the outputs of generator and discriminator are given as a feedback for fine-tuning as shown in Fig. 2(b).

The generator architecture as shown in Fig. 3 is a modified U-net architecture. The network is initialized with a downsampling block (DB) that consists of convolutional layer, batchnorm layer for regularization and leaky rectified linear unit (LReLU). The convolutional layer increases the number of channels to 64, uses the filters with size 4×4 with stride 2. After each DB, the output size reduces by a factor of two in both lateral dimensions. Similarly, other convolutional blocks were serially added to maximize the number of channels. Further, a series of up-sampling blocks (UB) was used in the network in a reverse manner to reduce the number of channels and maintain the original size. Each upsampling block contains transposed convolutional layer followed by batchnorm, dropout and ReLU activation. At the input of each UB, the output from preceding UB is concatenated with the output of DB at the same level (see Fig. 3. A convolutional layer in the end matches the size of the label by reducing the number of output channels.

The discriminator is a PatchGAN architecture[31], consists of sequential layers of five downsampling blocks, two zero padding layers followed by a final convolutional layer with a sigmoid function. Every 70x70 portion of the input image is classified by a 30x30 patch of the output. It represents the probability of the input being either real or fake and is used in calculating the loss functions for both, generator and discriminator separately.

The discriminator loss ($l_{disc}$) and the generator loss ($l_{gen}$) is calculated using sigmoid cross entropy as a function of discriminator outputs, $D(HR)$ and $D(G(LR))$. The discriminator and generator loss function are defined by equations (14) and (15) respectively.

$$l_{disc} = -(\log D(HR) + (\log(1 - D(G(LR))))) \quad (14)$$

$$l_{gen} = -(\log D(G(LR))) \quad (15)$$

The total generator loss ($l_{total\_gen}$) is the composite function of sigmoid cross entropy - a function of $D(G(LR))$ and mean absolute error (MAE) between real image and the generated image. The MAE is used as L1 loss that regularizes the generator model to predict images that are a plausible translation of the ground truth.

$$MAE = |HR - G(LR)| \quad (16)$$

$$l_{total\_gen} = -(\log D(G(LR)) + \alpha |HR - G(LR)|) \quad (17)$$

After calculating the losses, the trainable parameters were updated using an adaptive moment estimation (Adam) optimizer with a learning rate 2×10⁻⁴ for both generator and discriminator networks. The hyperparameter $\alpha$ was set to 100 after optimizing with multiple trials. The random normal initializer was used to initiate the convolution layers in first downsampling and upsampling blocks. The truncated normal distribution was used to initialize weights while zero initializer was adopted for the network bias terms. The RBC and macrophages training datasets contained 2355 and 2279 images while the testing datasets had 505 and 210 images respectively, each of size 512×512 pixels. A batch size of 1, input image size 512×512 and the buffer size of 500 were used during the training of the network. The model was trained for 50 epochs separately for RBC and macrophages datasets which took ~11 hrs for each training loop. The GAN losses and MAE were logged into the tensorboard for epoch optimization. The diminishing trend of MAE can be observed in Fig. 3. Hence the scaled L1 loss or MAE is added to $l_{gen}$ to calculate $l_{total\_gen}$ before applying the gradients. As the MAE decreases with epochs (see Fig. 3) it is evident that the matching between generated images and the ground truth has improved.

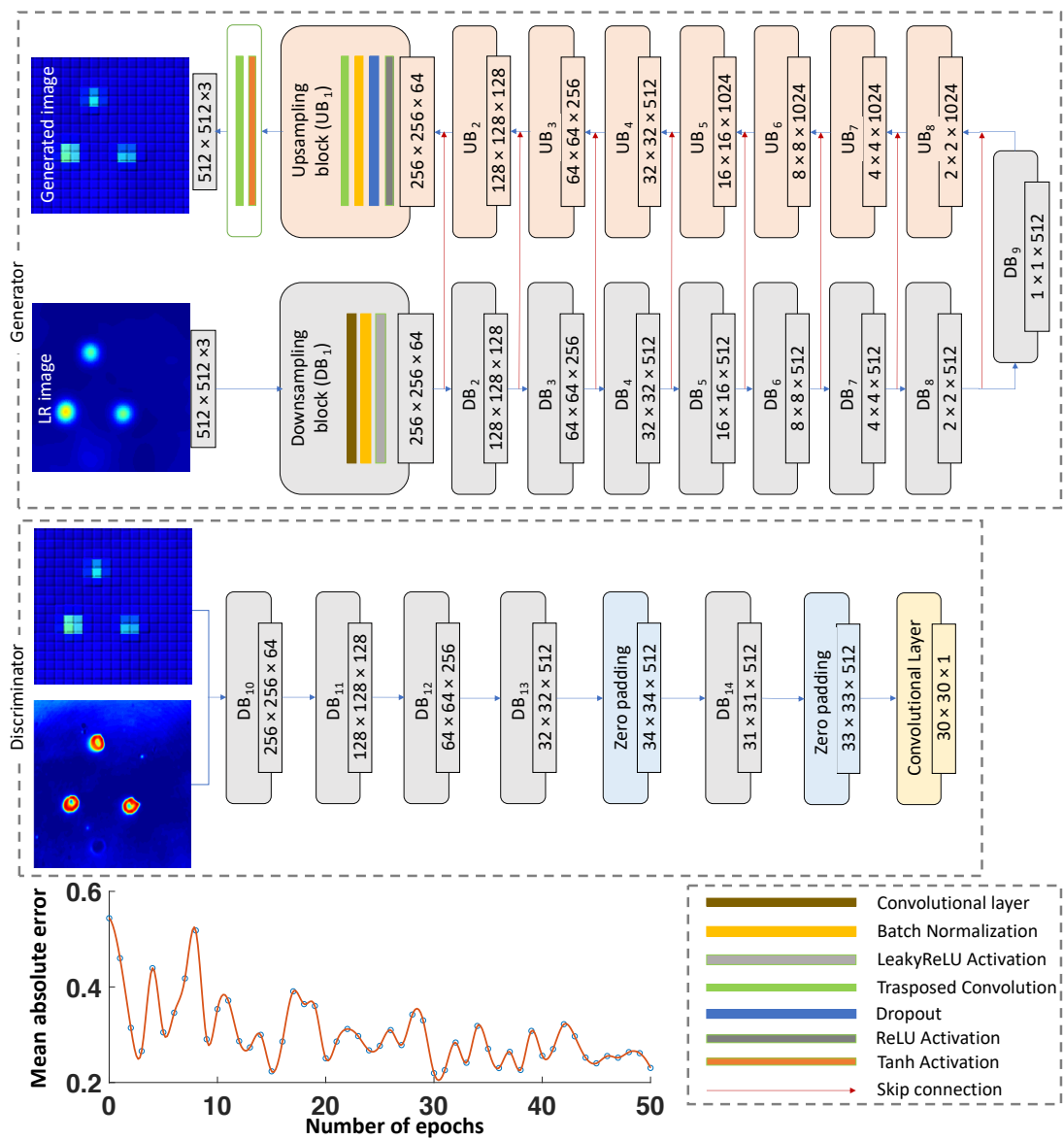

Figure 3: Architecture of generative adversarial network to generate high resolution phase image: Structure of generator and discriminator of the network. The mean absolute error (MAE) decreases with number of epochs indicates the matching between generated phase images and the high resolution i.e. ground truth phase image.

## 3. Results and discussion

The results first show the comparison of reconstructed phase from the experimental datasets and the phase prediction for RBC and macrophages cell lines from the network. In addition, the framework is scalable for two different datasets where two different resolution are achieved. To predict HR images from the LR phase, we first optimize and train the network for RBC datasets and without any hyperparameter tuning it was trained separately to generate HR macrophages phase images. Further, unseen images provided to the trained network and compared against the ground truth. Specifically, Fig. 4 shows the comparison of the LR, HR (ground truth) and the predicted images of the RBC and macrophages. Since LR image is captured with low NA lens therefore, high frequency components are missing in the network input image. In other words, less number of "k-vectors" of spatial frequencies are captured by the low NA lens. Additionally, line profile along the dashed line shows that donut shapes are not visible in LR image (Fig 4a). The line profile along the same pixel are shown for both HR image and predicted image. The HR phase map and performance of the neural network output can be seen from Fig. 4 (b) and (c). The line profile of both images shows the variation of phase map along the same pixels. Phase map at the membrane of the RBC varies between 1-2.5 rad for the HR image and matching approximately with the network predicted phase. The sphericity and shape of different RBC can be estimated from the network output.

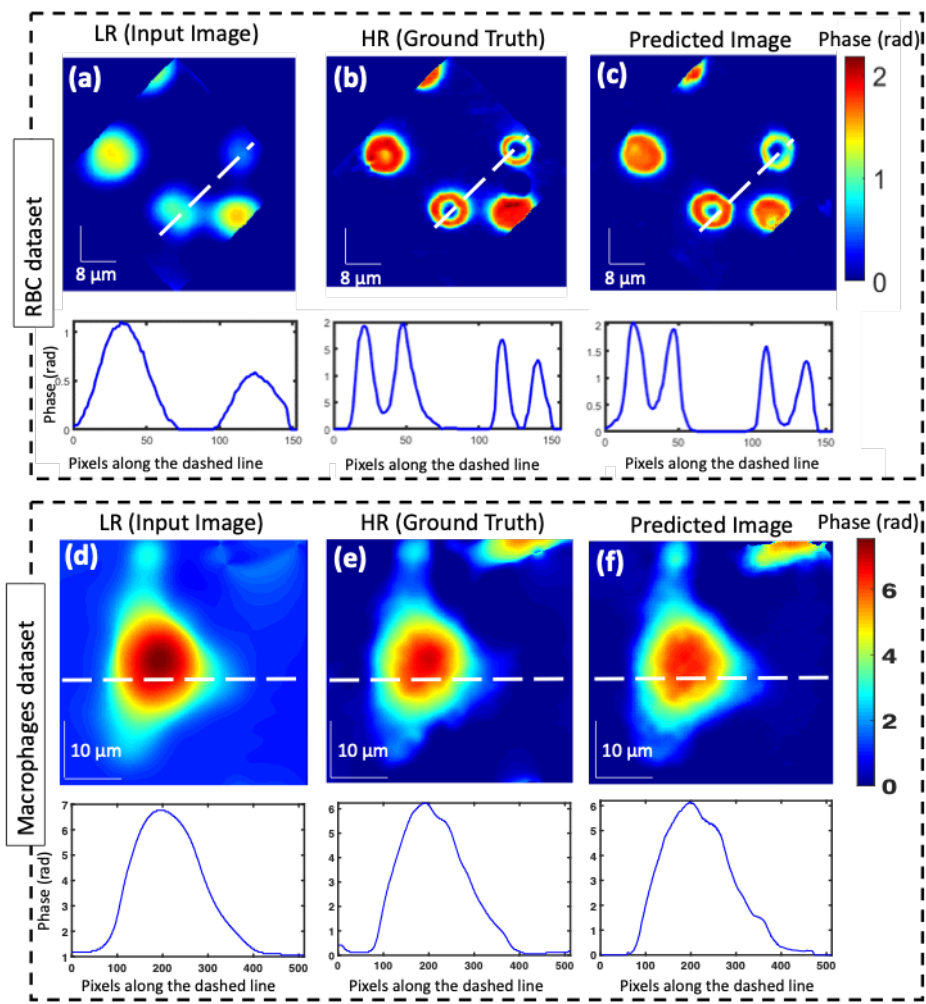

Figure 4: High-resolution phase estimation for human red blood cell (RBC) and macrophages from PSC-OCM+DNN framework. (a) The low resolution (LR) and (b) high resolution (HR) phase reconstructed from the experimental setup. The LR image taken as an input and (c) predicted HR by the network. Similar comparison (d) LR, (e) HR and (f) predicted phase of macrophages datasets is shown here. The line profile along the same pixel for both datasets shows the comparison of predicted phase map with the input image and ground truth. The LR and HR datasets of RBC

is acquired by 10X, 0.25 NA and 50X, 0.75 NA respectively. For macrophages, the LR images acquired by 20X, 0.45 NA and HR images using 60X, 1.2 NA objective lens. Color bar shows phase map in radian.

Similarly, the comparison between LR, HR and predicted phase is shown for macrophages dataset in Fig. 4(d-f). The macrophages are more complex structure than the RBC, thus critically estimating the robustness of our framework. Unlike the RBC datasets, the macrophages are heterogenous with its shape and size varies largely in the spatial dimension. The phase value of macrophages varies from 2 rad to 10 rad with different cell structures hence more robust training is required to predict the HR datasets. The line profile of the HR and predicted macrophages can be seen in Fig. 4(e) and (f). We observe that phase value of the whole structure including nucleus and membrane of the macrophages was correctly predicted by the network output. However, some mismatch found in the predicted images, which can be occurred and difficult to avoid in practice, because of the experimental imperfections such as illumination, spatially varying aberration due to optical components, sample preparations and phase unwrapping artifacts. The comparison and losses in the spatial frequency between the network output and ground truth images are shown by measuring structured similarity index (SSIM) [32] and spatial frequency spectrum.

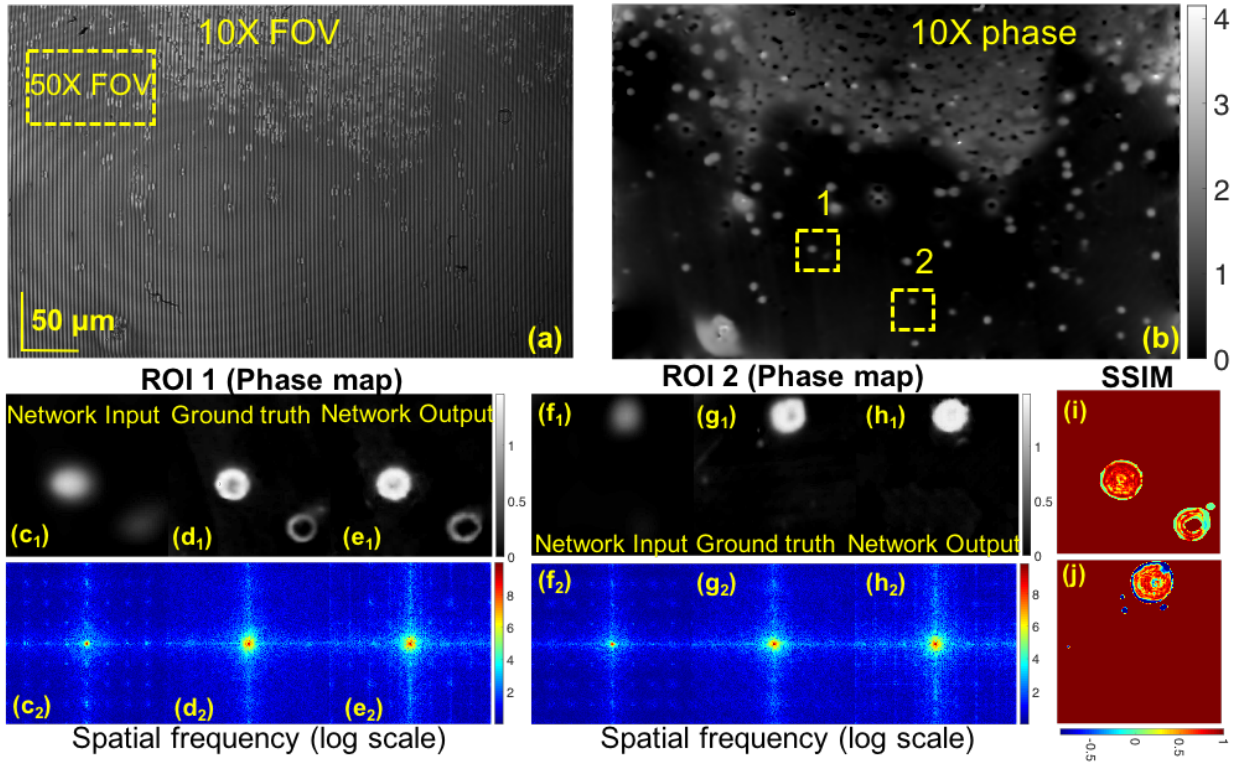

Figure 5: Multiple ROI phase prediction for human RBC datasets. (a, b): The full-FOV of 10X and reconstructed phase map are shown here. The small ROI 1 and 2 zoom-in and compare with the ground truth (high resolution phase) and network output (predicted phase). The broadening in spatial frequency spectrum in the predicted phase of ROI 1 and ROI 2 shows performance of framework to achieving higher frequency component. (i, j) represents the structured similarity index (SSIM) between ground truth and predicted phase of ROI 1 and 2, respectively. SSIM quantify the differences between network output and ground truth image[32]. SSIM index for ROI 1 and ROI 2 is found to be 0.96 and 0.97, respectively.

In Fig. 5, we present the comparison of full field of view (FOV) and predicted phase for RBC datasets. The interferogram and reconstructed phase map for 10X NA is shown in Fig. 5 (a, b). Since FT based phase reconstruction algorithm does not utilize the full resolution of the system, therefore less information is visible in the whole FOV phase image. We show the performance of our architecture to produce HR images for multiple ROI in Fig. 5. The ROI 1 and 2 are upsampled (Fig. 5 ($c_1$) and ($f_1$)) up to same pixel size as HR image and compared with the ground truth and predicted phase map. The predicted phase map of ROI 1 (Fig. 5 ($e_1$)) and ROI 2 (Fig. 5 ($h_1$)) shows notable resolution enhancement in the LR image and matched approximately with the ground truth i.e., Fig. 5 ($d_1$) and ($g_1$).

Further, resolution enhancement in the predicted image is shown by spatial frequency analysis. Figure 5 ($c_2$-$e_2$) and ($f_2$-$h_2$) shows the spatial frequency spectrum for LR, HR and predicted images for ROI 1 and 2, respectively. The broadening in spatial frequency (log scale) spectrum is another indication of the performance of the network to achieve higher frequency components. The spatial frequency spectrum of the predicted images is matching closer to the ground truth image. The difference in structural information and spatial frequency can be quantified by measuring SSIM index [32]. Fig. 5 (i) and (j) depicts the SSIM between ground truth and predicted phase of ROI 1 and 2, respectively. The SSIM index varies between -1 and 1 where 1 can achieve if predicted and ground truth images are identical to each other. In our case, the SSIM index for ROI 1 and ROI 2 is found to be 0.96 and 0.97, respectively. These values show high similarity between ground truth and network output images for the RBC datasets. However, the effect of optical components, illumination and phase related artifacts cannot be neglected and hence affect the network prediction and the spatial frequency spectrum of the predicted images.

Figure 6 shows multiple ROI comparison between network generated and ground truth phase map for macrophages datasets. The bright field, interferogram and reconstructed phase map for 20X NA is depicted in Fig. 6 (a-c). Similar to the RBC datasets, the performance of the network generated phase map for two different ROI are compared with the ground truth phase images. The predicted phase image compares with the network input and the ground truth both in spatial and frequency domain. The broadening in spatial frequency spectrum in the predicted phase of ROI 1 and frequency spectrum along $f_x$ in network output for ROI 2 shows performance of framework to achieving higher frequency component. The SSIM index for ROI 1 and ROI 2 is found to be 0.79 and 0.87, respectively. The SSIM index of ROI 1 is found less and can be understood by seeing the phase unwrapping artefact in the ground truth image and no artefact observed in the predicted image. Additionally, the mismatch between HR and predicted images can be seen at the edges of the cells since it consists the high frequency components of the specimens. Nevertheless, the predicted images show approximate matching between ground truth and predicted phase images for macrophages datasets.

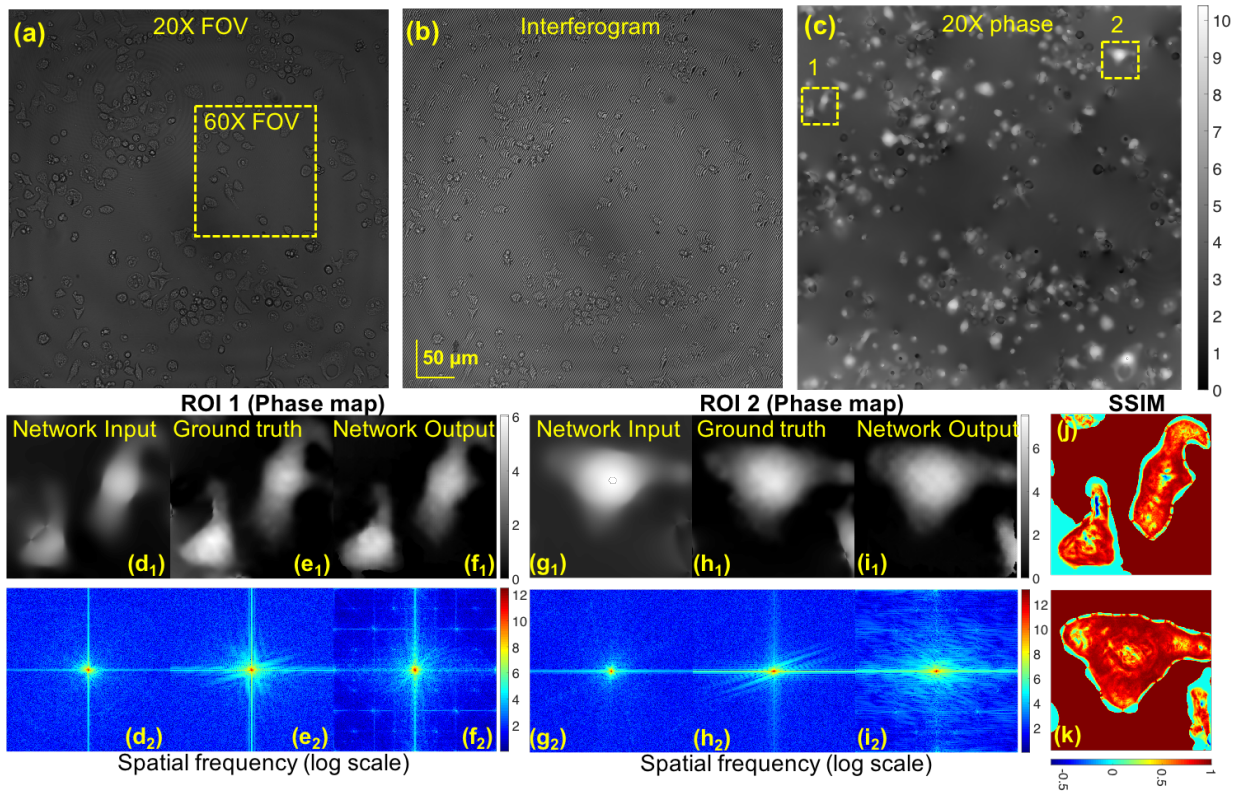

Figure 6: Multiple ROI phase prediction for macrophages using PSC-OCM+DNN framework. (a-c): The full-FOV of 20X, experimentally recorded interferogram and reconstructed phase map. ($d_1$-$f_1$), ($g_1$-$f_1$): The small ROI 1 and 2 zoom-in and compare with the ground truth (high resolution

phase) and network output (predicted phase). (d$_2$-f$_2$), (g$_2$-f$_2$): The network output compared by measuring spatial frequency spectrum. The broadening in spatial frequency spectrum in the predicted phase of ROI 1 and frequency spectrum along $f_x$ in network output for ROI 2 shows performance of framework to achieving higher frequency component. (j, k): The difference in structural information is compared with calculating structural similarity index for ROI 1 and 2, respectively. SSIM index for ROI 1 and ROI 2 is found to be 0.79 and 0.87, respectively. Phase unwrapping artefact in the ground truth image results poor value of SSIM for ROI 1.

However, careful examination must be required to interpret the high-resolution phase prediction of PSC-OCM+DNN framework. For instance, although the LR and HR phase images are mapped and resolution enhancement is clearly visible, the fine information of the cells (both macrophages and RBC) are slightly mismatched. Further, other artefacts due to source illumination, spatial phase sensitivity, temporal phase sensitivity and phase reconstruction algorithm cannot be completely avoidable. Since the spatial phase sensitivity of the PSC-OCM system is ±20 mrad thus any phase information equal or smaller than ±20 mrad likely represents the artefact. Also, temporal phase sensitivity represents the temporal stability the system thus any displacement even in the nanometer range might create the mismatch between HR and predicted phase map. The additional complications can be caused by the phase unwrapping algorithm. For example, Goldstein et al.[33] and minimum Lp norm method [34] generate streaky artefacts in the unwrapped phase image especially in case of thick sample such as macrophages. However, the artefacts are less in TIE based unwrapping which does not require any prefiltering and perform faster than the other methods. Though the artefacts cannot be completely neglected and therefore create problem in the matching between the predicted and ground truth image. Furthermore, the sample preparations and sub-pixel matching of upsampled LR and HR increases the challenges to predict HR phase image. Nonetheless, the proposed framework produced promising HR phase map and will be useful for incoherent phase imaging technique and other label-free imaging technique where averaging the speckle pattern can improve the resolution thus offer artefact free imaging.

4. Conclusion

We have presented a partially spatially coherent gated optical coherence microscopy (PSC-OCM) system assisted with generative adversarial network (GAN) for high SBP imaging of different biological specimens. The PSC system was synthesized by introducing the spatial and temporal diversity in the path of direct monochromatic laser source. Spatial and temporal diversity are introduced to reduce the spatial coherence of the laser source and thus for speckle free imaging purpose. The speckle free interferometric images are acquired and further reconstructed by FT and TIE algorithm to extract the phase map of the human RBC and macrophages cells lines. Further, the reconstructed low-resolution phase map is used to train and optimize the GAN architecture. The trained network is used to produce the high-resolution phase map of the specimen. The performance of the network was demonstrated on both human RBC and macrophages cell lines. The GAN predicted phase image further compared with the ground truth in both spatial and Fourier domain. The SSIM index value shows good matching between ground truth and predicted phase images. It is shown that the framework offers label free, single-shot platform for high-resolution imaging of the specimens, without any further requirement of optimization and hyperparameter tuning.

Our proposed high spatially sensitive, high-resolution phase imaging framework that can support large FOV will find usage in applications that require imaging over large areas such as whole tissue slide imaging, neuronal movements [35]. Moreover, the proposed method will also be useful in applications that require high-speed imaging, such as QPI of sperms cells [36], and QPI of trapped RBC [37, 38]. The current approach can further be used in longitudinal spatial coherence (LSC) gated full-field optical coherence tomography (FF-OCT) system for high resolution tomography of the biological specimens by utilizing LSC properties of the source. Additionally, the proposed approach will also find application in material sciences, for example for profilometry of devices, optical waveguides, silicon circuits, etc benefitting of imaging over larger areas with improved resolution and sensitivity.

**References:**

1. G. Popescu, *Quantitative phase imaging of cells and tissues* (McGraw Hill Professional, 2011).
2. E. Cuche, F. Bevilacqua, and C. Depeursinge, "Digital holography for quantitative phase-contrast imaging," Optics letters **24**, 291-293 (1999).
3. G. Popescu, T. Ikeda, R. R. Dasari, and M. S. Feld, "Diffraction phase microscopy for quantifying cell structure and dynamics," Optics letters **31**, 775-777 (2006).

4. A. Butola, D. Popova, D. K. Prasad, A. Ahmad, A. Habib, J. C. Tinguely, P. Basnet, G. Acharya, P. Senthilkumaran, and D. S. Mehta, "High spatially sensitive quantitative phase imaging assisted with deep neural network for classification of human spermatozoa under stressed condition," arXiv preprint arXiv:.07377 (2020).
5. G. Zheng, R. Horstmeyer, and C. Yang, "Wide-field, high-resolution Fourier ptychographic microscopy," Nature photonics **7**, 739 (2013).
6. V. P. Pandiyan, K. Khare, and R. John, "Quantitative phase imaging of live cells with near on-axis digital holographic microscopy using constrained optimization approach," Journal of biomedical optics **21**, 106003 (2016).
7. M. Trusiak, V. Mico, J. Garcia, and K. Patorski, "Quantitative phase imaging by single-shot Hilbert–Huang phase microscopy," Optics letters **41**, 4344-4347 (2016).
8. Y. Xue, S. Cheng, Y. Li, and L. Tian, "Reliable deep-learning-based phase imaging with uncertainty quantification," Optica **6**, 618-629 (2019).
9. T. Liu, K. de Haan, Y. Rivenson, Z. Wei, X. Zeng, Y. Zhang, and A. Ozcan, "Deep learning-based super-resolution in coherent imaging systems," Scientific reports **9**, 1-13 (2019).
10. Y. Jo, S. Park, J. Jung, J. Yoon, H. Joo, M.-h. Kim, S.-J. Kang, M. C. Choi, S. Y. Lee, and Y. Park, "Holographic deep learning for rapid optical screening of anthrax spores," Science advances **3**, e1700606 (2017).
11. A. S. Singh, A. Anand, R. A. Leitgeb, and B. Javidi, "Lateral shearing digital holographic imaging of small biological specimens," Optics express **20**, 23617-23622 (2012).
12. Y. Rivenson, Y. Zhang, H. Günaydın, D. Teng, and A. Ozcan, "Phase recovery and holographic image reconstruction using deep learning in neural networks," Light: Science Applications **7**, 17141-17141 (2018).
13. T. Nguyen, Y. Xue, Y. Li, L. Tian, and G. Nehmetallah, "Deep learning approach for Fourier ptychography microscopy," Optics express **26**, 26470-26484 (2018).
14. C. L. Chen, A. Mahjoubfar, L.-C. Tai, I. K. Blaby, A. Huang, K. R. Niazi, and B. Jalali, "Deep learning in label-free cell classification," Scientific reports **6**, 21471 (2016).
15. Y. N. Nygate, M. Levi, S. K. Mirsky, N. A. Turko, M. Rubin, I. Barnea, G. Dardikman-Yoffe, M. Haifler, A. Shalev, and N. T. Shaked, "Holographic virtual staining of individual biological cells," Proceedings of the National Academy of Sciences **117**, 9223-9231 (2020).
16. Y. N. Nygate, G. Singh, I. Barnea, and N. T. Shaked, "Simultaneous off-axis multiplexed holography and regular fluorescence microscopy of biological cells," Optics letters **43**, 2587-2590 (2018).
17. T. Kim, R. Zhou, M. Mir, S. D. Babacan, P. S. Carney, L. L. Goddard, and G. Popescu, "White-light diffraction tomography of unlabelled live cells," Nature Photonics **8**, 256 (2014).
18. T. H. Nguyen, M. E. Kandel, M. Rubessa, M. B. Wheeler, and G. Popescu, "Gradient light interference microscopy for 3D imaging of unlabeled specimens," Nature communications **8**, 1-9 (2017).
19. J. Rosen, A. Vijayakumar, M. Kumar, M. R. Rai, R. Kelner, Y. Kashter, A. Bulbul, and S. Mukherjee, "Recent advances in self-interference incoherent digital holography," Advances in Optics Photonics **11**, 1-66 (2019).
20. J. Rosen and M. Takeda, "Longitudinal spatial coherence applied for surface profilometry," Applied optics **39**, 4107-4111 (2000).
21. M. Gokhler, Z. Duan, J. Rosen, and M. Takeda, "Spatial coherence radar applied for tilted surface profilometry," Optical Engineering **42**, 830-837 (2003).
22. I. Abdulhalim, "Spatial and temporal coherence effects in interference microscopy and full-field optical coherence tomography," Annalen der Physik **524**, 787-804 (2012).
23. A. Safrani and I. Abdulhalim, "Spatial coherence effect on layer thickness determination in narrowband full-field optical coherence tomography," Applied optics **50**, 3021-3027 (2011).
24. D. N. Naik, T. Ezawa, Y. Miyamoto, and M. Takeda, "Phase-shift coherence holography," Optics letters **35**, 1728-1730 (2010).
25. J. Heil, H.-M. Heuck, W. Müller, M. Netsch, and J. Wesner, "Interferometric spatial coherence tomography: focusing fringe contrast to planes of interest using a quasi-monochromatic structured light source," Applied optics **51**, 3059-3070 (2012).
26. A. Ahmad, T. Mahanty, V. Dubey, A. Butola, B. S. Ahluwalia, and D. S. Mehta, "Effect on the longitudinal coherence properties of a pseudothermal light source as a function of source size and temporal coherence," Optics letters **44**, 1817-1820 (2019).
27. D. S. Mehta, D. N. Naik, R. K. Singh, and M. Takeda, "Laser speckle reduction by multimode optical fiber bundle with combined temporal, spatial, and angular diversity," Applied optics **51**, 1894-1904 (2012).

28. M. Takeda, H. Ina, and S. Kobayashi, "Fourier-transform method of fringe-pattern analysis for computer-based topography and interferometry," JosA **72**, 156-160 (1982).
29. D. Paganin and K. A. Nugent, "Noninterferometric phase imaging with partially coherent light," Physical review letters **80**, 2586 (1998).
30. N. Pandey, A. Ghosh, and K. Khare, "Two-dimensional phase unwrapping using the transport of intensity equation," Applied optics **55**, 2418-2425 (2016).
31. P. Isola, J.-Y. Zhu, T. Zhou, and A. A. Efros, "Image-to-image translation with conditional adversarial networks," in *Proceedings of the IEEE conference on computer vision and pattern recognition*, 2017), 1125-1134.
32. Z. Wang, A. C. Bovik, H. R. Sheikh, and E. P. Simoncelli, "Image quality assessment: from error visibility to structural similarity," IEEE transactions on image processing **13**, 600-612 (2004).
33. R. M. Goldstein, H. A. Zebker, and C. L. Werner, "Satellite radar interferometry: Two-dimensional phase unwrapping," Radio science **23**, 713-720 (1988).
34. D. C. Ghiglia and L. A. Romero, "Minimum Lp-norm two-dimensional phase unwrapping," JOSA A **13**, 1999-2013 (1996).
35. K. C. Boyle, T. Ling, V. Zuckerman, T. Flores, and D. V. Palanker, "Quantitative phase imaging of neuronal movement during action potential (Conference Presentation)," in *Quantitative Phase Imaging VI*, (International Society for Optics and Photonics, 2020), 112490S.
36. A. Butola, D. Popova, A. Ahmad, V. Dubey, G. Acharya, P. Banet, P. Senthilkumaran, B. S. Ahluwalia, and D. S. Mehta, "Classification of human spermatozoa using quantitative phase imaging and machine learning," in *Digital Holography and Three-Dimensional Imaging*, (Optical Society of America, 2019), Th4A. 3.
37. A. Ahmad, V. Dubey, V. R. Singh, J.-C. Tinguely, C. I. Øie, D. L. Wolfson, D. S. Mehta, P. T. So, and B. S. Ahluwalia, "Quantitative phase microscopy of red blood cells during planar trapping and propulsion," Lab on a Chip **18**, 3025-3036 (2018).
38. B. S. Ahluwalia, P. McCourt, A. Oteiza, J. S. Wilkinson, T. R. Huser, and O. G. Hellesø, "Squeezing red blood cells on an optical waveguide to monitor cell deformability during blood storage," Analyst **140**, 223-229 (2015).